\documentclass{aa}
\usepackage[utf8]{inputenc}
 
\usepackage{graphicx,ulem}
\usepackage{longtable}
\usepackage{booktabs}
\usepackage{float}
\usepackage{dcolumn}
\usepackage{graphics,epsfig}
\usepackage{amsmath,amssymb,latexsym,mathrsfs}
\usepackage[breaklinks, colorlinks, citecolor=blue]{hyperref}
\usepackage{bm}
\usepackage{color}
\usepackage{subfigure}

\newcommand\horsp{\rule[-2mm]{0mm}{6mm}}
\newcommand\morehorsp{\rule[-3mm]{0mm}{8mm}}
\newcommand\mmorehorsp{\rule[-5mm]{0mm}{10mm}}

\usepackage{etoolbox}
\makeatletter
\patchcmd\@combinedblfloats{\box\@outputbox}{\unvbox\@outputbox}{}{%
   \errmessage{\noexpand\@combinedblfloats could not be patched}%
}%
 \makeatother

\begin{document}

\title{Constraints from thermal Sunyaev-Zeldovich cluster counts and power spectrum combined with CMB}
\titlerunning{Joint analysis of CMB, tSZ cluster counts and power spectrum}

\author{Laura Salvati\inst{1} \and Marian Douspis\inst{1} \and Nabila Aghanim\inst{1}}
\institute{Institut d'Astrophysique Spatiale, CNRS (UMR 8617) Université Paris-Sud, Bâtiment 121, Orsay, France
\email{laura.salvati@ias.u-psud.fr}}

\authorrunning{Salvati et al.}

\date{}

\abstract{
The thermal Sunyaev-Zel'dovich (tSZ) effect is one of the recent probes of cosmology and large-scale structures. We update constraints on cosmological parameters from galaxy clusters observed by the Planck satellite in a first attempt to combine cluster number counts and the power spectrum of hot gas; we used a new value of the optical depth and, at the same time, sampling  on cosmological and scaling-relation parameters. We find that in the $\Lambda$CDM model, the addition of a tSZ power spectrum provides small improvements with respect to number counts alone, leading to the $68\%$ c.l. constraints $\Omega _m = 0.32 \pm 0.02$, $\sigma _8 = 0.76\pm0.03, $ and $\sigma _8 (\Omega _m/0.3)^{1/3}= 0.78\pm0.03$ and lowering the discrepancy with results for cosmic microwave background (CMB) primary anisotropies (updated with the new value of $\tau$) to $\simeq 1.8\, \sigma$ on $\sigma _8$. We analysed extensions to the standard model, considering the effect of massive neutrinos and varying the equation of state parameter for dark energy. In the first case, we find that the addition of the tSZ power spectrum helps in improving cosmological constraints with respect to number count alone results, leading to the $95\%$ upper limit $\sum m_{\nu}< 1.88 \, \text{eV}$. For the varying dark energy equation of state scenario, we find no important improvements when adding tSZ power spectrum, but still the combination of tSZ probes is able to provide constraints, producing $w = -1.0\pm 0.2$. In all cosmological scenarios, the mass bias to reconcile CMB and tSZ probes remains low at $(1-b)\lesssim 0.67$ as compared to estimates from weak lensing and X-ray mass estimate comparisons or numerical simulations.
}


\maketitle

\section{Introduction}\label{sec:intro}

Galaxy clusters are the most massive bound structures emerging in the cosmic web of large-scale structure (LSS). These objects are associated with peaks in the matter density field on megaparsec scales. The abundance of clusters is strongly sensitive to both growth of structure and matter density, depends on the underlying cosmological model, and thus provides constraints on cosmological parameters; see e.g. \cite{Allen:2011zs}.

These constraints are even more powerful when combined with, or compared to, results from other observables; in particular primary temperature and polarization anisotropies of the cosmic microwave background (CMB) radiation and baryon acoustic oscillations (BAO). On the one hand, comparing results from various observables provides important consistency checks. On the other hand, the combination of geometrical and growth-based probes can improve constraints on parameters such as the equation of state (EoS) for dark energy owing to the different degeneracies between parameters for the different probes. 

Galaxy clusters are made by dark matter and baryons in different phases that can be probed by observations at different wavelengths; see again \cite{Allen:2011zs}. In recent years, several measurements of cluster samples in the X-rays \citep{Boehringer:2017wvr,Chon:2011gp}, optical \citep{Rykoff:2016trm}, and millimetre (mm)
\citep[][South Pole Telescope; SPT]{Bleem:2014iim}, \citep[][Atacama Cosmology Telescope; ACT]{Marriage:2010cp}, \citep[][Planck]{2014A&A...571A..29P,Ade:2015gva} wavelengths have improved the constraints on cosmological parameters. 

In this work, we focus on galaxy clusters observed in mm wavelengths through the thermal Sunyaev-Zel'dovich (tSZ) effect \citep{Sunyaev:1970er}, that is the inverse Compton scattering between CMB photons and hot electrons in the intra-cluster medium (ICM) using measurements of the Planck satellite  \citep{Adam:2015rua,Ade:2015gva}. In particular, we exploit the combination of galaxy cluster number counts and the angular power spectrum of warm-hot gas seen in SZ by Planck \citep{2016A&A...594A..22P} and SPT \citep{George:2014oba}. We try to quantify if and how the addition of current tSZ power spectrum data helps to better break the degeneracy between the cosmological parameters and those used to model the physics of the clusters.

In light of the discrepancy between CMB  and number counts constraints \citep{2014A&A...571A..20P}, we compare our results with most recent CMB data from \cite{2016A&A...594A..11P, Aghanim:2016yuo} for the $\Lambda$CDM model. We also explore results obtained by relaxing some assumptions of the standard model, in particular considering the sum of the neutrino masses, $\sum m_{\nu}$, and the dark energy EoS parameter, $w$, as varying parameters. We show how our combined analysis improves constraints on these extensions of the standard model.

This paper is organized as follows: In section \ref{sec:theory} we briefly describe the theoretical model needed to build the number counts for galaxy clusters observed through the tSZ effect and the model for the tSZ power spectrum. In section \ref{sec:method} we describe the approach we use in this analysis and in section \ref{sec:results} we show our results. In sections \ref{sec:discussion} and \ref{sec:conclusions} we derive our final discussion and conclusions.

\section{tSZ cosmological probes}\label{sec:theory}

The thermal Sunyaev-Zel'dovich effect is a powerful cosmological probe. The main property of this effect is the fact that its surface brightness is redshift independent, therefore providing nearly mass-limited cluster samples from high-resolution mm surveys at arbitrarily high redshift. The ability to sample up to high redshifts ensures that we can track with high accuracy the evolution of LSS; in particular this enables us to constrain neutrino mass owing to its effect on the evolution of LSS, i.e. the damping of matter power spectrum at small scales; see e.g. \cite{Lesgourgues:2012uu}.

The intensity of the tSZ effect, in a given direction of the sky $\hat{\bf n}$, is measured through the thermal Compton parameter $y$, defined as
\begin{equation}\label{eq:y}
y(\hat{\bf n}) = \int n_{\text{e}} \dfrac{k_{\text{B}}T_{\text{e}}}{m_{\text{e}} c^2} \, \sigma_{\text{T}} \, ds \, ,
\end{equation}

\noindent where $k_{\text{B}}$ is the Boltzmann constant, $\sigma_{\text{T}}$ is the Thomson scattering cross section, and $m_{\text{e}}$, $n_{\text{e}}$, and $T_{\text{e}}$ are the electron mass, number density, and temperature, respectively. 

To define clusters detected through the tSZ effect, we adopt the following convention: the cluster mass $M_{500}$ is the total mass contained in a sphere of radius $R_{500}$, which is defined as the radius within which the cluster mean mass overdensity is 500 times the critical density at that redshift, i.e.
\begin{equation}\label{eq:M500}
M_{500}= \dfrac{4 \pi}{3} R_{500}^3 500 \rho _{\text{c}} (z) \, ,
\end{equation}
where the critical density is defined as
\begin{equation}\label{eq:rho_cr}
\rho _{\text{c}} (z) = \dfrac{3H^2(z)}{8 \pi G}
\end{equation}
and $H(z)$ is the Hubble parameter. 

We consider therefore the following observables for clusters detection from \cite{Ade:2015gva}: $Y_{500}$, which is the Compton $y$-profile integrated within a sphere of radius, $R_{500}$, and the cluster angular size, $\theta _{500}$.

\subsection{Number counts}\label{sec:NC}

The predicted number of clusters observed by a given survey in a redshift bin $[z_i,z_{i+1}]$ is given by
\begin{equation}\label{eq:counts1}
n_i = \int _{z_i} ^{z_{i+1}} dz \dfrac{dN}{dz} \, ,
\end{equation}
\noindent where
\begin{equation}\label{eq:counts2}
\dfrac{dN}{dz} = \int d\Omega \int _{M_{\text{min}}} ^{M_{\text{max}}}  dM_{500} \, \hat{\chi} (z,M_{500};l,b) \, \dfrac{dN}{dz \, dM_{500} \, d \Omega} \, ,
\end{equation}
\noindent where $\hat{\chi} (z,M_{500};l,b)$ is the survey completeness at a given position in the sky $(l,b)$ and 
\begin{equation}\label{eq:counts3}
\dfrac{dN}{dz \, dM_{500} \, d \Omega} = \dfrac{dN(M_{500},z)}{dM_{500}} \dfrac{dV_c}{dz \, d\Omega}
\end{equation}
\noindent is the product between the comoving volume element (per unit redshift and solid angle) $dV_c/dz d\Omega$ and the mass function $dN(M_{500},z)/dM_{500}$. The latter represents the probability of having a galaxy cluster of mass $M$ at redshift $z$, per unit volume, in the direction given by $d\Omega$.

This description can be generalized to define number counts as functions of the signal-to-noise ratio as well; see \citep{Ade:2015fva}.

\subsection{tSZ power spectrum}\label{sec:tSZ}

The complete analytical description of tSZ power spectrum has been fully covered in different papers, such as \cite{Komatsu:2002wc} and \cite{2014A&A...571A..21P}; therefore we only report the necessary results.

Considering the halo-model (see e.g. \cite{Cooray:2000ge}), we can write tSZ power spectrum as the sum of one-halo and two-halo terms,
\begin{equation}\label{eq:Cell}
C_{\ell}^{\text{tSZ}} = C_{\ell} ^{1\, \text{halo}} + C_{\ell} ^{2\, \text{halo}} \, .
\end{equation}

In the flat sky limit ($\ell \gg 1$), the one-halo term is expressed as
\begin{eqnarray}\label{eq:Cell1halo}
C_{\ell} ^{1\, \text{halo}} &=& \int _0 ^{z_{\text{max}}} dz \dfrac{dV_c}{dz\, d\Omega} \notag \\
& \times & \int _{M_{\text{min}}} ^{M_{\text{max}}} dM \dfrac{dN(M_{500},z)}{dM_{500}} |\tilde{y}_{\ell}(M_{500},z)| ^2 \notag \\
& \times & \exp{\left( \dfrac{1}{2} \sigma ^2 _{\ln Y^*}  \right)} \, .
\end{eqnarray}

\noindent The term $\tilde{y}_{\ell}(M_{500},z)$ is the Fourier transform on the sphere of the Compton parameter $y$ of individual clusters and is given by (using Limber approximation)\begin{eqnarray}\label{eq:y_tilde}
\tilde{y}_{\ell}(M_{500},z) &=&  \dfrac{4 \pi r_{\text{s}}}{\ell ^2 _{\text{s}}}  \left( \dfrac{\sigma _{\text{T}}}{m_{\text{e}} c^2} \right)  \notag \\
& \times & \int _0 ^{\infty} dx \, x^2 \, P_{\text{e}} (M_{500},z,x) \dfrac{\sin (\ell x/\ell _s)}{\ell x/\ell _s} \, ,
\end{eqnarray}

\noindent where $r _{\text{s}}$ is the scale radius of the 3D pressure profile, $P_{\text{e}} (M_{500},z,x)$, $\ell _{\text{s}} = D_A(z)/r _{\text{s}}$ (where $D_A(z)$ is the angular diameter distance) and $x = r/r _{\text{s}}$. We use the universal pressure profile provided by \cite{Arnaud:2009tt}. The term $\sigma _{\ln Y^*}$ represents the dispersion in the scaling relations that is fully described in the next section. 

The two-halo term \citep{Komatsu:1999ev} is derived from computing the correlation between two different halos as
\begin{eqnarray}\label{eq:Cell2halo}
C_{\ell} ^{2\, \text{halo}} &=& \int _0 ^{z_{\text{max}}} dz  \dfrac{dV_c}{dz\, d\Omega}   \notag \\
 & \times & \left[   \int  _{M_{\text{min}}} ^{M_{\text{max}}}  dM \dfrac{dN(M_{500},z)}{dM_{500}} \tilde{y}_{\ell}(M_{500},z)  B(M_{500},z) \right] ^2 \notag \\
 & \times & P(k,z) \,, 
\end{eqnarray}

\noindent where $P(k,z)$ is the matter power spectrum and $B(M,z)$ is the time-dependent linear bias factor. This last term relates the matter power spectrum to the power spectrum of the cluster correlation function. We follow \cite{Komatsu:1999ev} and use the definition
\begin{equation}\label{eq:beta_func}
B(M_{500},z)= 1 + \dfrac{\nu ^2 (M_{500},z)}{\delta _c (z)} \, ,
\end{equation}

\noindent where $\nu(M_{500},z) = \delta _c(M_{500})/D(z)\sigma(M_{500})$and  $\sigma(M_{500})$ is the present-day rms mass fluctuation, $D(z)$ the linear growth factor, and $\delta _c (z)$ the threshold over-density of spherical collapse.

Following the analysis of \cite{Komatsu:2002wc} and \cite{Horowitz:2016dwk}, to evaluate our errors accurately, we also take into account the contribution from the trispectrum term, $T_{\ell \ell'}$, which is the harmonic-space four-point function and represents the non-Gaussian contribution of the cosmic variance. The dominant term in the halo model is defined as \citep{Cooray:2001wa,Komatsu:2002wc}
\begin{eqnarray}\label{eq:T_ll}
T_{\ell \ell'} & \simeq & \int _0 ^{z_{\text{max}}} dz \dfrac{dV_c}{dz d\Omega} \notag \\
& \times & \int _{M_{\text{min}}} ^{M_{\text{max}}} dM \left[ \dfrac{dN(M_{500},z)}{dM_{500}} \right. \notag \\
& \times & |\tilde{y}_{\ell}(M_{500},z)|^2 |\tilde{y}_{\ell'}(M_{500},z)|^2 \biggr] \, .
\end{eqnarray}

\subsection{Assumptions on the modelling ingredients}

There are some major uncertainties in constraining cosmological parameters from galaxy clusters that are related to the ingredients needed to build the theoretical model described in the previous section. In fact the theoretical mass function needs to be transformed into a prediction of the distribution of clusters in the parameter space of survey observables. The first uncertainty is therefore related to an imperfect knowledge of the scaling relations between the cluster mass and the survey observable, that is used as a proxy for this quantity. The second uncertainty is related to the model of the selection process, which needs to be accurately built, to avoid possible selection biases that could affect final results on cosmological parameters constraints.

Thus,  we must make some assumptions on scaling relations, mass function, and selection function to completely determine the theoretical model. Regarding the latter, it is out of the scope of this paper, we therefore refer to the complete discussion reported in \cite{Ade:2015fva}.

\subsubsection{Scaling relations}
A crucial element in modelling the cosmological probes is the exact evaluation of clusters mass and therefore of the scaling relations between survey observables and mass. For all details on the evaluation of these relations for the Planck observables, we refer to \cite{2014A&A...571A..20P} and we report only the final formulas for the integrated Compton $y$-profile, $Y_{500}$,
\begin{equation}\label{eq:Y500}
E^{-\beta} (z) \left[ \dfrac{D_A^2 (z) Y_{500}}{10^{-4} \, \text{Mpc}^2} \right] = Y_* \left[ \dfrac{h}{0.7}\right] ^{-2 + \alpha} \left[ \dfrac{(1-b) M_{500}}{6 \cdot 10^{14} M_{\odot}} \right] ^{\alpha}
\end{equation}
and for the cluster angular size
\begin{equation}\label{eq:theta500}
\theta_{500} = \theta _* \left[ \dfrac{h}{0.7}\right] ^{-2/3} \left[ \dfrac{(1-b) M_{500}}{3 \cdot 10^{14} M_{\odot}} \right]^{1/3} E^{-2/3} (z) \left[ \dfrac{D_A(z)}{500 \, \text{Mpc}}\right] ^{-1} \, .
\end{equation}
\noindent In equations \eqref{eq:Y500} and \eqref{eq:theta500} $D_A(z)$ is again the angular diameter distance, $h$ is the reduced Hubble constant, $H_0/100$, and $E(z)=H(z)/H_0$. For the coefficients we follow what is reported in \cite{Ade:2015fva} and consider $\theta_*= 6.997 \, \text{arcmin}$, $\beta=0.66$ and for the others coefficients we use Gaussian distributed priors, reported in table \ref{tab:SR_priors}. Equation \eqref{eq:Y500}  is derived with a dispersion, $\sigma _{\ln Y_*}$, given in table \ref{tab:SR_priors}. We consider it as a nuisance parameter for the counts only, since its effect on the power spectrum amplitude is negligible (lower than $1\%$). Therefore we neglect the last term in Eq. \eqref{eq:Cell1halo}. 

The quantity $b$ is defined as the mass bias and it takes into account the difference between the cluster mass estimation, obtained assuming hydrostatic equilibrium, and the real cluster mass. By following the \cite{Ade:2015fva} baseline, in this analysis we use a Gaussian distributed prior for $(1-b)$, from the Canadian Cluster Comparison Project \citep[][labelled CCCP form now on]{Hoekstra:2015gda}, reported in table \ref{tab:SR_priors}. In section \ref{sec:discussion}, we discuss in detail the status of measurements and constraints on this quantity and its effects on the final results on cosmological parameters.

\begin{table}[!h]
\begin{center}
\scalebox{1.0}{
\begin{tabular}{cc}
\hline
\hline
\horsp
 Parameter & Gaussian prior \\
\hline
\morehorsp
 $\log Y_* $ & $-0.19 \pm 0.02$\\
\hline
\morehorsp
$\alpha$ & $1.79 \pm 0.08$\\
\hline
\morehorsp
 $\sigma _{\ln Y_*}$ & $0.173 \pm 0.023$ \\
\hline
\morehorsp
 $(1-b)$ CCCP & $0.780 \pm 0.092$\\
\hline
\end{tabular}}
\caption{\footnotesize{Priors on nuisance parameters for scaling relations, as defined in \cite{Ade:2015fva}.}}
\label{tab:SR_priors}
\end{center}
\end{table}

\subsubsection{Mass function}
In order to evaluate the theoretical mass function, we rely on numerical $N$-body simulations. In particular, for our analysis we use the mass function provided by \cite{Tinker:2008ff}. Therefore the number of halos per unit volume is given by
\begin{equation}\label{eq:Tinker}
\dfrac{dN}{dM_{500}} = f(\sigma) \dfrac{\rho _{\text{m},0}}{M_{500}}\,  \dfrac{d \ln \sigma ^{-1}}{dM_{500}} \, ,
\end{equation}
where $\rho _{\text{m},0}$ is the matter density at redshift $z=0$ and
\begin{equation}\label{eq:f_sigma}
f(\sigma) = A \left[ 1 + \left( \dfrac{\sigma}{b}\right) ^{-a} \right] \exp{\left( -\dfrac{c}{\sigma ^2}\right)} \, .
\end{equation}
In Eqs. \eqref{eq:Tinker} and \eqref{eq:f_sigma}, $\sigma$ is the standard deviation of density perturbations in a sphere of radius $R=(3M/4 \pi \rho _{\text{m},0})^{1/3}$, calculated in linear regime, and it is given by
\begin{equation}\label{eq:sigma}
\sigma ^2 = \dfrac{1}{2 \pi ^2} \int dk \, k^2 P(k,z) \, |W(kR)|^2 \, ,
\end{equation}
\noindent where $W(kR)$ is the window function of a spherical top hat of radius $R$.

We evaluated the coefficients $A$, $a$, $b,$ and $c$ in eq. \eqref{eq:f_sigma} by interpolating the results provided by \cite{Tinker:2008ff} at the required overdensity (i.e. $500 \, \rho_{\text{c}} (z)$).

Different $N$-body simulations, but also different analyses (e.g. halo finder methods) produce various fitting formulas for the mass function. \cite{Ade:2015fva} compared results on the cosmological parameters obtained with mass functions from \cite{Tinker:2008ff} and \cite{2013MNRAS.433.1230W}. Changing mass function does not change the accuracy in constraining cosmological parameters and it produces only a shift of $\sim 1 \, \sigma$ on the final constraints. 
We choose here to use the mass function from \cite{Tinker:2008ff} since this is the most used in cluster analyses and since it offers direct comparison with \cite{Ade:2015fva} for which this mass function was the baseline.

\section{Method}\label{sec:method}
In this work, we constrained the cosmological parameters from galaxy clusters, exploiting the combination of number counts and power spectrum. We considered the cluster sample provided by \cite{Ade:2015gva}, which consists of 439 clusters from the 65\% cleanest part of the sky, above the signal-to-noise ratio threshold of 6 and in the redshift range $z=[0,1]$. In order to obtain the cluster number counts, we sampled on both redshift and signal-to-noise ratio bins, as described in \cite{Ade:2015fva}. For the power spectrum, we used Planck estimates from \cite{2016A&A...594A..22P} and an estimate of the angular power spectrum from SPT at $\ell = 3000$ \citep{George:2014oba}. We integrated in the redshift range $z=[0,3]$ and in the mass range $M_{500} = [10^{13} h^{-1}M_{\odot},5 \cdot 10^{15} h^{-1}M_{\odot}]$, following \cite{2016A&A...594A..22P}.  
In combining cluster number counts and tSZ power spectrum, we followed the analysis shown in \cite{Hurier:2017jgi}, who have found a low level of correlation between cluster number counts and tSZ power spectrum. This is due to varying contributions to the variance for the two probes, depending on the mass range. In particular, for the tSZ power spectrum the main contribution comes from massive halos ($M_{500}>10^5 M_{\odot}$), while for the number counts the main contribution comes from lower mass halos. This small overlap between the two galaxy cluster populations results in a small correlation. Therefore in our combination of tSZ cluster number counts and power spectrum, we decided to neglect any correlation between the two tSZ probes

We used a Monte Carlo Markov Chains (MCMC) approach to sample and constrain at the same time cosmological and scaling-relation parameters, which we consider as nuisance parameters in this analysis. When considering tSZ power spectrum data from Planck, we used the error bars already marginalized over the foreground and noise contributions (in particular clustered cosmic infrared background, radio point sources, infrared point sources, and correlated noise), as described in \cite{2016A&A...594A..22P}. We stress that even if SZ number counts and power spectrum show similar dependencies on cosmological parameters, they have different dependencies on scaling relations parameter $\alpha$. We find $dN/dz \propto \sigma_8 ^9 \, \Omega_m ^3 (1-b)^{3.6}$ and $C_{\ell}^{\text{tSZ}} \propto \sigma _8^{8.1} \, \Omega _m^{3.2} (1-b)^{3.2}$. Combining the two probes, we should therefore be able to reduce the degeneracy between nuisance and cosmological parameters. 

For the present analysis, we used the November 2016 version of the publicly available package \texttt{cosmomc} \citep{Lewis:2002ah}, relying on a convergence diagnostic based on Gelman and Rubin statistics. This version includes the cluster number-count likelihood for Planck \citep{Ade:2015fva}. We modified this version to add the likelihood for the tSZ power spectrum \citep{2014A&A...571A..21P}. 

For the cosmological model, we first considered the $\Lambda$CDM model, varying the six standard parameters: the baryon and cold dark matter densities, $\omega_b$ and $\omega_c$;  ratio of the sound horizon to the angular diameter distance at decoupling $\theta$; scalar spectral index, $n_s$; overall normalization of the spectrum, $A_s$ at $k = 0.05 \, \text{Mpc}^{-1}$; and reionization optical depth $\tau$. We updated the results from \cite{Ade:2015xua} with the new value of the optical depth from \cite{Aghanim:2016yuo,Adam:2016hgk} by adding a Gaussian prior in our analysis, i.e. $\tau = 0.055 \pm 0.009$. We included in our analysis the four scaling parameters reported in table \ref{tab:SR_priors}, i.e. $Y_*$, $\alpha$, $(1-b)$ and $\sigma _{\ln Y_*}$, only for number counts. Since SZ number counts and power spectrum are not able to constrain the basic six-parameters model alone, we also added BAO measurements from \cite{Anderson:2013zyy}. We compared and combined our results with CMB primary temperature and polarisation anisotropy data from \cite{Ade:2015xua};  we updated these results with the new optical depth reported above.

Finally, we explored results obtained by relaxing the assumptions of the standard model, i.e. allowing first the sum of neutrino masses $\sum m_{\nu}$ and then the dark energy EoS parameter $w$ to vary, and therefore added these to our analysis.

\section{Results}\label{sec:results}
We report here our results, comparing constraints from tSZ power spectrum alone, tSZ number counts alone, and their combination. We compared these results with those obtained from CMB temperature and polarization anisotropy data and the complete combination of datasets (power spectrum, number counts, and CMB data). We stress that when considering tSZ probes, alone or in combination with CMB data, we always add BAO measurements as well. We analysed both the standard $\Lambda$CDM model and extensions to it. 

We present results for cosmological parameters to which galaxy clusters are more sensitive, in particular the total matter density, $\Omega _m$, and the standard deviation of density perturbations, defined in Eq. \eqref{eq:sigma}, evaluated at radius $R=8 \, \text{Mpc} \, h^{-1}$, $\sigma _8$.

\subsection{$\Lambda \text{CDM}$ model}
\begin{figure}[!ht]
\centering
\includegraphics[scale=0.19]{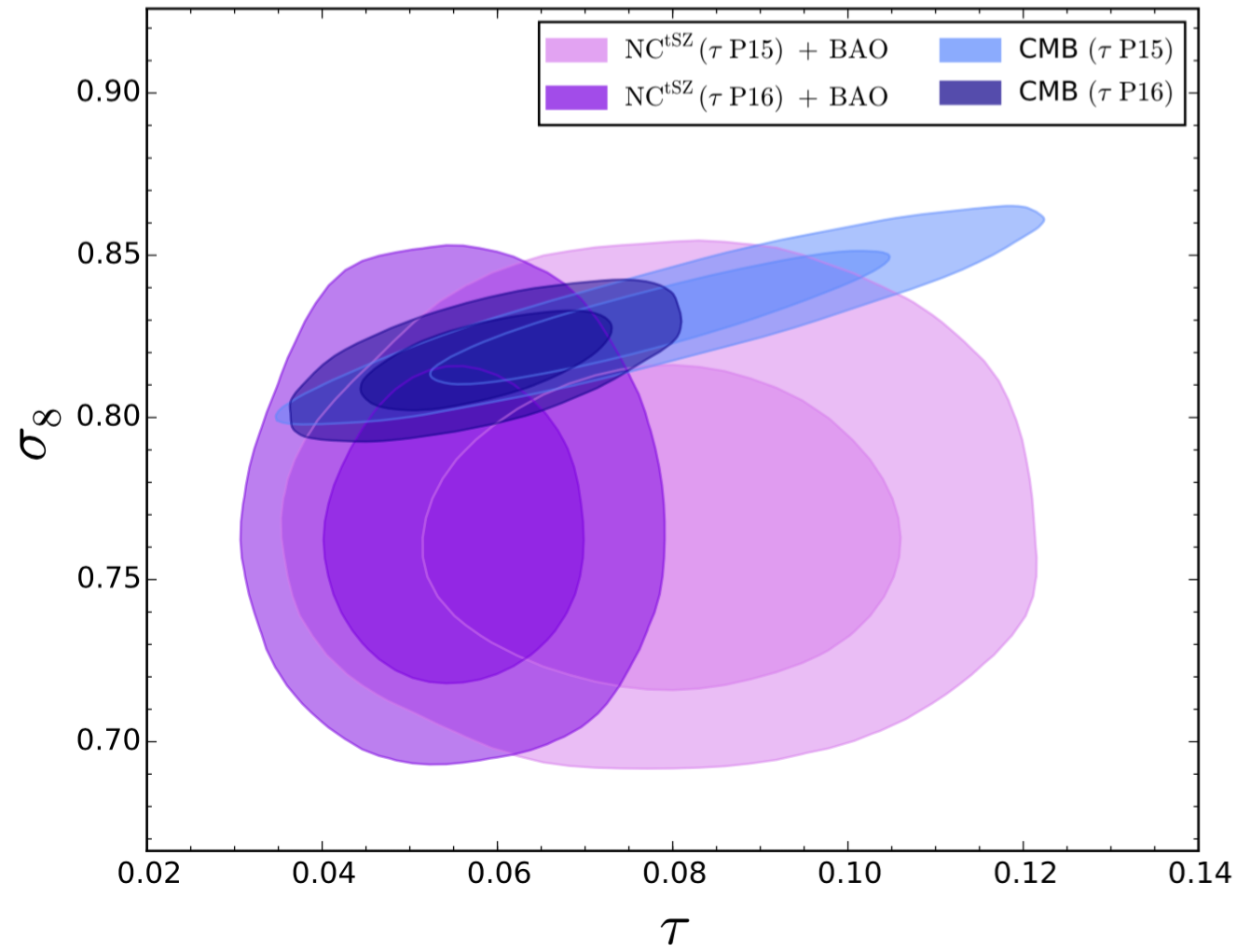}
\caption{\footnotesize{Two-dimensional probability distributions for $\tau$ and $\sigma _8$ for various values of optical depths (see text). We compare results for SZ number counts alone (pink and purple) and for CMB data alone (blue and light blue).}}
\label{fig:LCDM_tau_comparison}
\end{figure}

We first show the effect of the new value of the optical depth. In Fig. \ref{fig:LCDM_tau_comparison}, we compare two-dimensional probability distributions for $\tau$ and $\sigma _8$ for tSZ number counts and CMB data, for the various values of $\tau$. We find that while this change in the optical depth does not affect the constraining power of cluster number counts on $\sigma_8$, the change modifies constraints from CMB, therefore reducing the discrepancy between the two different probes. The change in CMB constraints is due to the degeneracy between optical depth and $\sigma _8$, that is the fact that small-scale CMB power spectrum is proportional to the quantity $\sigma _8 e^{-\tau}$  \cite[see e.g.][]{2014A&A...571A..16P}. The improved constraint on $\sigma _8$ from CMB reduces the tension with SZ number counts from $2.4\, \sigma $ \citep{Ade:2015xua} to $1.5 \, \sigma$. 

\begin{figure}[!ht]
\centering
\includegraphics[scale=0.33]{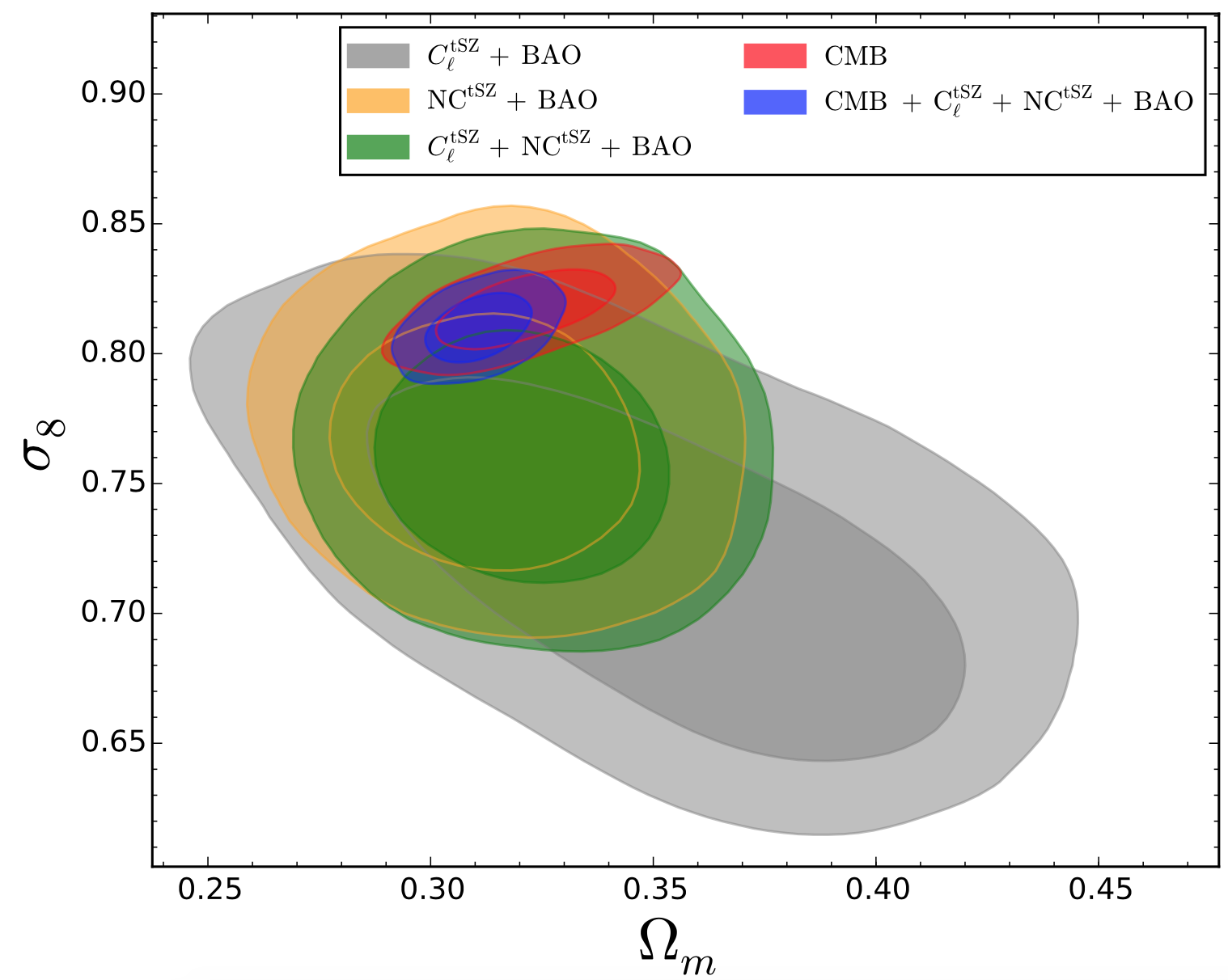}
\caption{\footnotesize{Two-dimensional probability distributions for $\Omega _m$ and $\sigma _8$ in the $\Lambda$CDM scenario, only power spectrum (grey), only number counts (orange), the combination of the two probes (green), only CMB (red), and the combination of all the probes (blue).}}
\label{fig:LCDM_sigma_omega}
\end{figure}

We focus now on the results for $\sigma _8$ and the matter density $\Omega _m$. We show the constraints from CMB temperature and polarization anisotropy data, from the tSZ power spectrum alone ($C_{\ell} ^{\text{tSZ}} $), from tSZ number counts alone ($\text{NC}^{\text{tSZ}}$), from the combination of the two tSZ probes ($C_{\ell} ^{\text{tSZ}} + \text{NC}^{\text{tSZ}}$), and finally those from the complete combination of all datasets ($\rm CMB + C_{\ell} ^{\text{tSZ}} + \text{NC}^{\text{tSZ}}$), adding BAO data as well. We stress that from now on we always use the new prior for the optical depth and that results for $C_{\ell} ^{\text{tSZ}}$ are always obtained by combining Planck and SPT data. We show the results in Fig. \ref{fig:LCDM_sigma_omega} and we summarize the constraints ($68\%$ c.l.) in Tab. \ref{tab:LCDM} for the various datasets.

\begin{figure*}[!ht]
\centering
\includegraphics[scale=0.6]{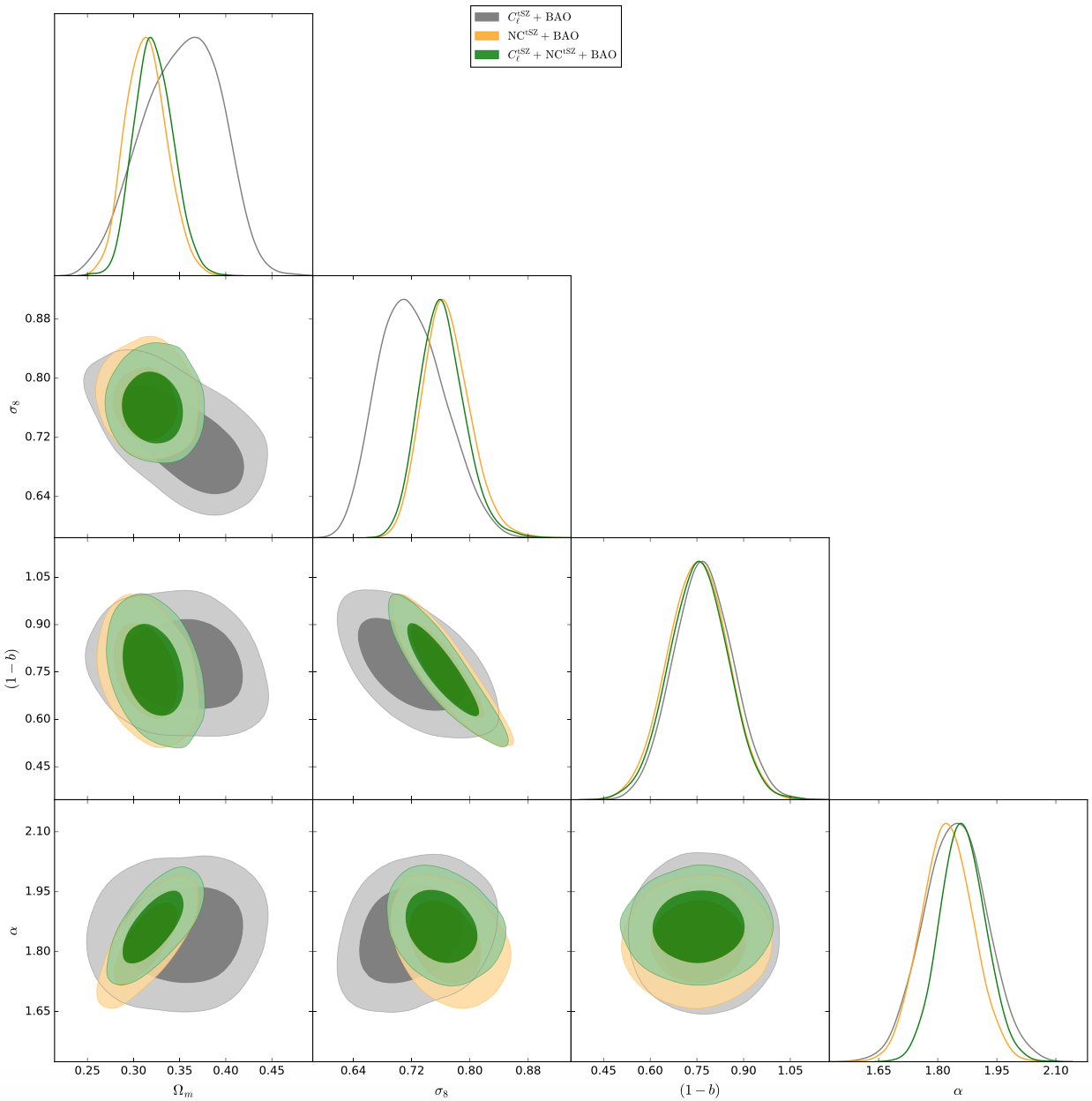}
\caption{\footnotesize{Correlation between cosmological and scaling-relation parameters in the case of tSZ power spectrum (grey), number counts (orange), and for the combination of both (green).}}
\label{fig:triangular}
\end{figure*}

When considering the combination of tSZ number counts and power spectrum, we note that the combination is driven by tSZ counts since tSZ spectrum shows weaker constraints; see comparison of the figure of merits (FoM) for the various datasets in Tab. \ref{tab:fom}. We nevertheless obtain a small improvement on the $\Omega_{\text{m}}$ and $\sigma _8$ constraints, within 10\% on individual error bars, and a small shift towards lower values of $\Omega _m$ and $\sigma _8$, within $0.2$ and $0.3 \, \sigma$. 
The slight differences in scaling-cosmological parameter degeneracies between the two tSZ probes drive this small improvement, as shown in Fig. \ref{fig:triangular}. As for the comparison between constraints from the CMB and tSZ combined probes, we find a slightly larger discrepancy, of $\simeq  \, 1.8 \, \sigma$, than the case of the CMB versus tSZ counts alone.

\begin{table}[!h]
\begin{center}
\scalebox{1.0}{
\begin{tabular}{cccc}
\hline
\hline
\morehorsp
 FoM & $C_{\ell}^{\text{tSZ}}$ & $\text{NC}^{\text{tSZ}}$ &$C_{\ell}^{\text{tSZ}}+\text{NC}^{\text{tSZ}}$ \\
\hline
\mmorehorsp
 $\dfrac{1}{\sigma _{\sigma _8}\sigma _{\Omega _m}}$ &$567$ &$1462$ & $1592$ \\
\hline
\end{tabular}}
\caption{\footnotesize{Figures of merit (FoM) for tSZ spectrum alone, number counts alone, the combination of the two probes,  and for $\Omega _m$ and $\sigma _8$ parameters in the $\Lambda$CDM scenario.}}
\label{tab:fom}
\end{center}
\end{table}

\begin{figure}[H]
\centering
\includegraphics[scale=0.37]{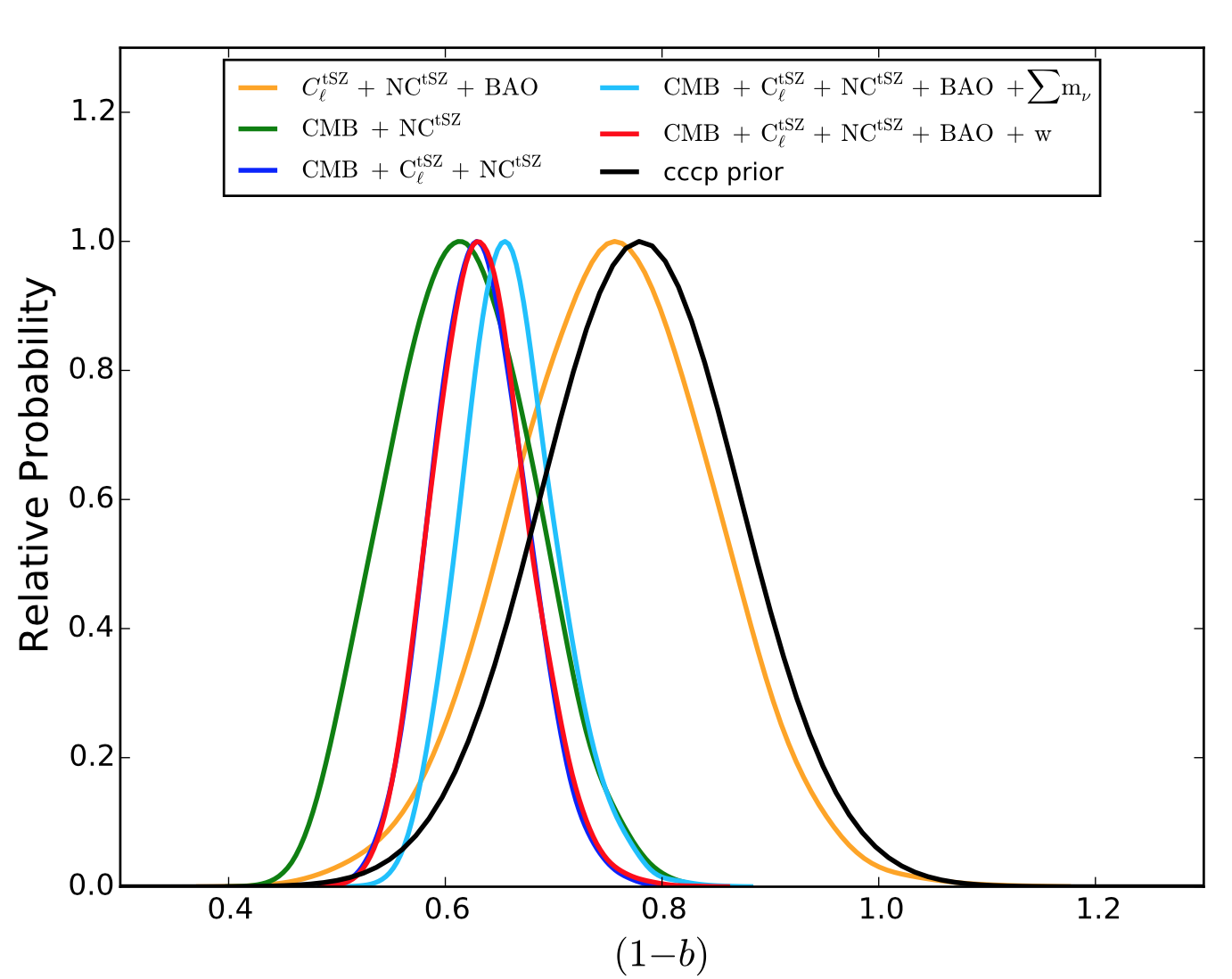}
\caption{\footnotesize{One-dimensional probability distribution for the mass bias $(1-b)$ for various dataset combinations: the complete tSZ combination and BAO (orange); CMB and the complete tSZ combination (blue, almost completely overlapped by the red line); CMB and number counts (green); the combination of CMB and tSZ, adding the effect of massive neutrinos (light blue); and the combination of CMB and tSZ, adding the effect of varying the dark energy EoS parameter (red). All of these combinations are compared to the CCCP prior we used in our analysis (black).}}
\label{fig:mass_bias}
\end{figure}

We now focus on the scaling-relation parameters and in particular on the mass bias, which significantly affects the values of $\sigma_8$. As noted in \cite{2014A&A...571A..20P,Ade:2015fva}, low values of mass bias lead to high values of $\sigma_8$ (see also Fig.~\ref{fig:triangular}). We show in Fig.~\ref{fig:mass_bias} the results from the tSZ combination probes,  adding the CMB data, together with the CCCP-based prior considered in our analysis.
In our updated analysis with the new optical depth, we find that results from the tSZ combined probes are driven by the prior distribution. Adding CMB data to the tSZ counts or to the combined tSZ probes drives the mass bias to lower values; in this case we do not add the BAO data in order to fully compare with results from \cite{Ade:2015fva}. On the one hand, we find that the bias needed to reconcile CMB constraints with those from the tSZ number counts is $(1-b) = 0.62 \pm 0.07$, which is comparable to the value $(1-b)=0.58 \pm 0.04$ found in  \cite{Ade:2015fva}. However, the bias increases to $(1-b)= 0.63 \pm 0.04$ when using the tSZ counts and power spectrum.

\subsection{Extensions to $\Lambda$CDM}

 We  now consider two extensions to the $\Lambda$CDM model: adding massive neutrinos and including the EoS of dark energy.
For these extensions to the standard model, we explore whether the combination of the two tSZ probes can  improve the constraints on cosmological and scaling-relation parameters with respect to only number counts. We compare  these various results with constraints from only CMB and from the complete combination of all datasets. 

\begin{table*}[!h]
\begin{center}
\scalebox{0.87}{
\begin{tabular}{c|c|c|c|c|c}
\hline
\hline
\horsp
 Cosmological parameters & $C_{\ell}^{\text{tSZ}}$+ BAO & $\text{NC}^{\text{tSZ}}$+ BAO & $C_{\ell}^{\text{tSZ}} + \text{NC}^{\text{tSZ}}$+ BAO & CMB & CMB + $C_{\ell}^{\text{tSZ}} + \text{NC}^{\text{tSZ}}$+ BAO  \\
\hline
\morehorsp
 $\Omega _{\text{m}}$ & $ 0.352 _{-0.038}^{+0.047} $ & $0.314_{-0.024}^{+0.020} $ & $0.322 _{-0.022}^{+0.020} $ & $ 0.321_{-0.014}^{+0.012} $ & $0.311 \pm 0.007 $\\
\hline
\morehorsp
$\sigma _8$ & $ 0.721 _{-0.053}^{+0.039} $ &$0.768_{-0.035}^{+0.028}$ & $ 0.762 _{-0.034}^{+0.027} $ & $ 0.817 \pm 0.010 $ & $0.810 \pm 0.008 $\\
\hline
\morehorsp
$S_8 = \sigma_8 (\Omega _m/0.3)^{1/3}$ & $ 0.759 _{-0.042}^{+0.035} $ &$ 0.780 _{-0.042}^{+0.029} $ & $0.780 _{-0.040}^{+0.028} $ & $ 0.836 \pm 0.018 $ & $0.820 \pm 0.012 $\\
\hline
\hline
\morehorsp
 Nuisance parameters & $C_{\ell}^{\text{tSZ}}$+ BAO & $\text{NC}^{\text{tSZ}}$+ BAO & $C_{\ell}^{\text{tSZ}} + \text{NC}^{\text{tSZ}}$+ BAO & CMB & CMB + $C_{\ell}^{\text{tSZ}} + \text{NC}^{\text{tSZ}}$+ BAO\\
 \hline
\morehorsp
$(1-b)$ & $ 0.770 \pm 0.092 $ & $0.754 \pm 0.093 $ & $ 0.755 \pm 0.091 $ &\bf - & $0.646 _{-0.039}^{+0.034} $\\
\hline
\morehorsp
 $\alpha$ & $ 1.845 \pm 0.077 $ & $1.824\pm 0.064 $ & $1.864 _{-0.060}^{+0.056} $ &\bf - & $1.822 _{-0.041}^{+0.036}$\\
\hline
\morehorsp
$\log Y_*$ & $-0.186 \pm 0.021 $ & $ -0.189\pm 0.020  $ & $ -0.189 \pm 0.020 $ &\bf - & $-0.194 \pm 0.021 $\\
\hline
\morehorsp
 $\sigma _{\ln Y_*}$ &\bf - & $0.075 \pm 0.010 $ & $ 0.075 \pm 0.010 $ &\bf - & $0.075 \pm 0.010 $\\
\hline 
\end{tabular}}
\caption{\footnotesize{$68\%$ c.l. constraints for cosmological and scaling-relation parameters in the $\Lambda$CDM scenario from power spectrum $(C_{\ell}^{\text{tSZ}})$ and number counts $(\text{NC}^{\text{tSZ}})$ alone and for the combination of the two probes $(C_{\ell}^{\text{tSZ}}+\text{NC}^{\text{tSZ}})$. We compare these results with constraints from CMB primary anisotropies and from the complete combination of datasets.}}
\label{tab:LCDM}
\end{center}
\end{table*}

\subsubsection{Massive neutrinos}
\begin{figure*}[!h]
  \centering
  \subfigure{\includegraphics[scale=0.24]{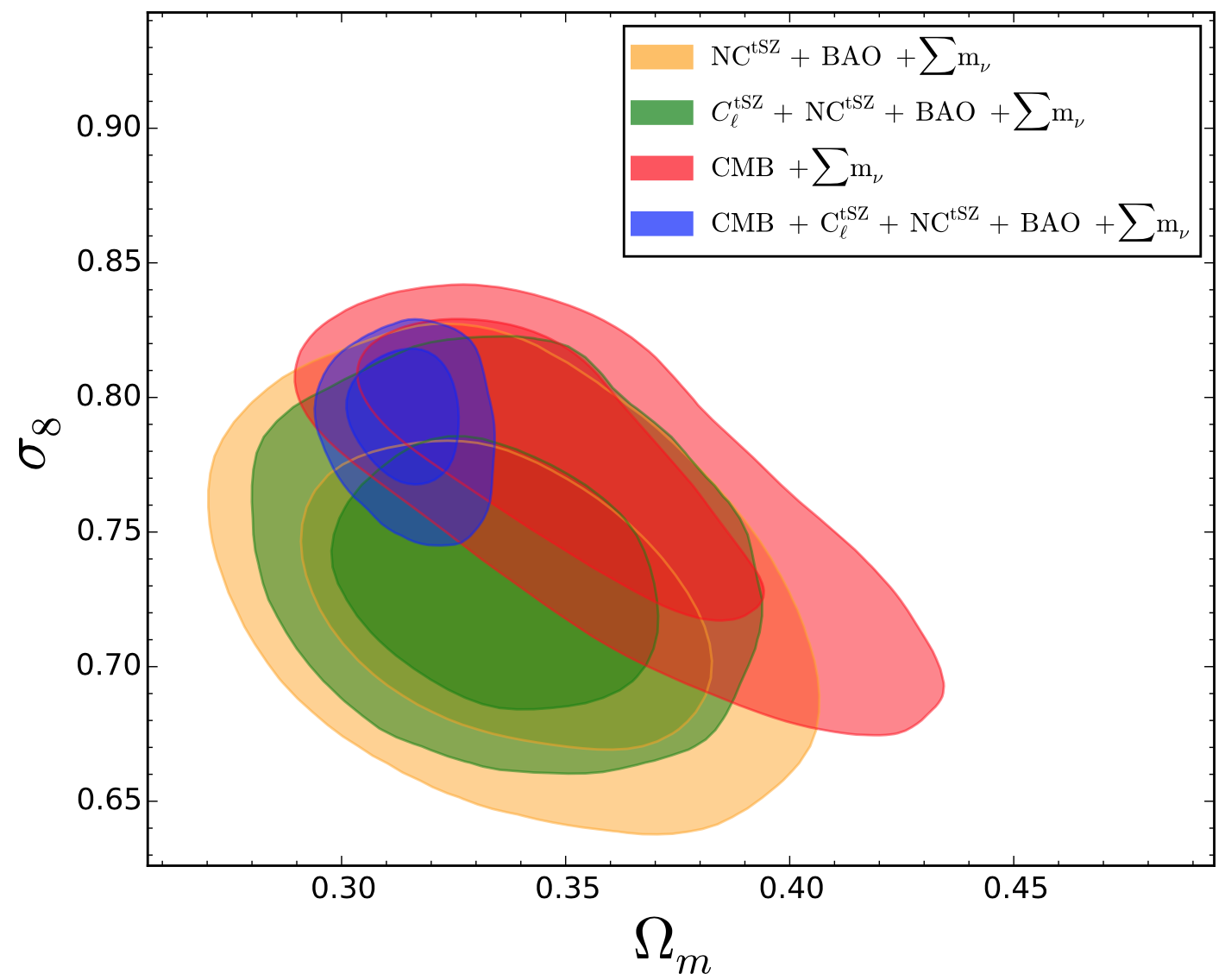}}\,
  \subfigure{\includegraphics[scale=0.242]{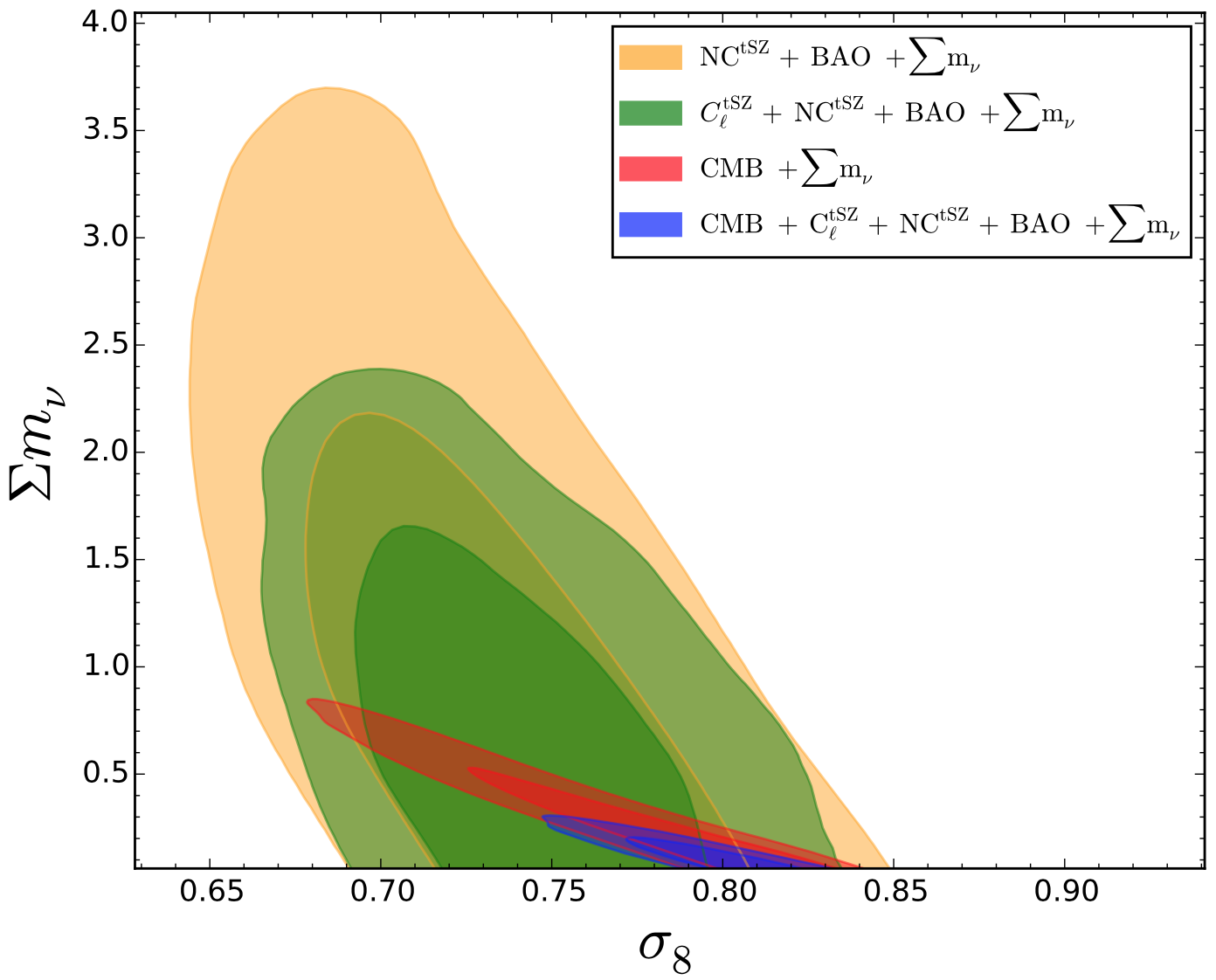}}\,
  \subfigure{\includegraphics[scale=0.228]{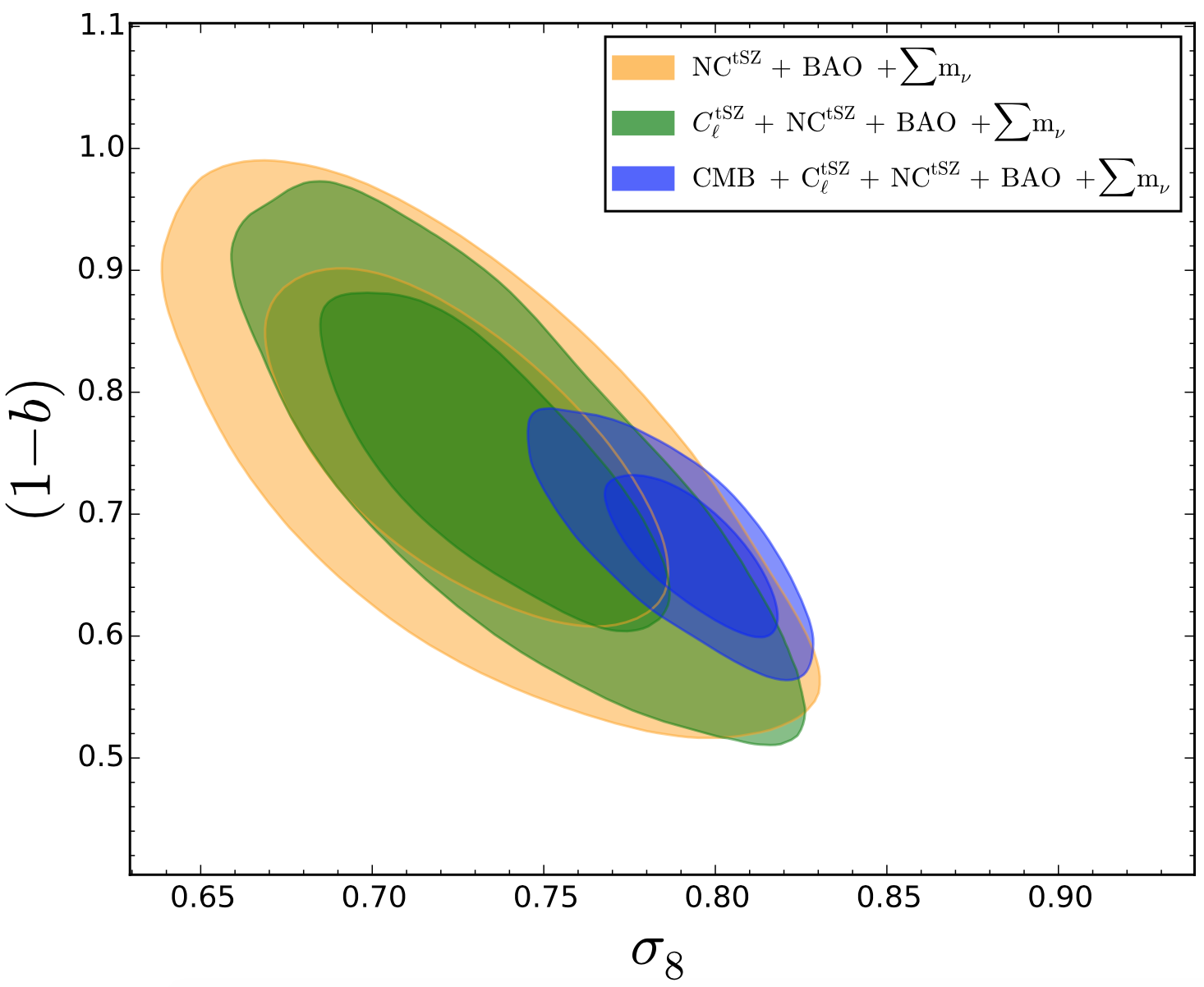}}
  \caption{\footnotesize{Two-dimensional probability distributions for $(\Omega _m,\sigma _8)$, $(\sigma_8,\sum m_{\nu}),$ and $(\sigma_8,(1-b))$ for varying neutrino mass scenario. We report results for number counts (orange), the combination of number counts and power spectrum (green); we add also CMB data (blue) and we show results for CMB alone (red).}}
  \label{fig:mnu_ALL_params}
\end{figure*}

For a given large-scale amplitude (constrained by CMB low multipoles), adding massive neutrinos damps the amplitude of matter power spectrum at small scales, which in turn lowers the value of $\sigma_8$. We present in the left panel of Fig. \ref{fig:mnu_ALL_params}  and in Tab. \ref{tab:mnu}, the new constraints on $\Omega_m$ and $\sigma_8$ obtained from number counts alone, CMB alone,  tSZ probes alone, and from the combination of the three. In performing the analysis with massive neutrinos, we exclude the values $\sum m_{\nu}<0.06 \, \text{eV}$, ruled out by neutrino oscillations experiments; see e.g. \cite{Patrignani:2016xqp}. Constraints from CMB primary anisotropies alone worsen, as compared to the $\Lambda$CDM case, because of the low sensitivity of CMB to the neutrino mass. As expected, lower values of $\sigma _8$ are reached, but along a $\Omega _m - \sigma _8$ degeneracy line, parallel to the tSZ degeneracy.

Constraints from tSZ probes alone are not significantly affected by the neutrino mass. As a matter of fact, the high S/N threshold of the Planck cluster sample selects massive clusters ($M \gtrsim 2 \times 10^{14}$) whose abundance is not impacted by matter power spectrum damping. The Planck tSZ power spectrum does not probe sufficiently well the small angular scales where the effect of the matter power spectrum damping due to massive neutrinos should take place. However, the addition of 
estimation of the tSZ power spectrum at $\ell=3000$ from SPT is expected to increase the sensitivity of the power spectrum to massive neutrinos. The full tSZ probe combination thus improves the final constraints on cosmological parameters with respect to number counts alone, as can be seen in Fig. \ref{fig:mnu_ALL_params} and in Tab. \ref{tab:mnu}. In particular, it provides an upper $95\%$ limit on neutrino mass $\sum m_{\nu} < 1.88 \, \text{eV}$, while number counts alone are only able to provide $\sum m_{\nu} < 2.84 \, \text{eV}$.

Despite the wider CMB constraints along the degeneracy line, we obtain an agreement within $1.3 \, \sigma$ between CMB and the tSZ probes. For the constraints obtained from the combination between tSZ probes (+BAO) and CMB, we stress that they are mainly driven by the latter, as can be seen in Fig. \ref{fig:mnu_ALL_params}. We show also the one-dimensional probability distribution for the mass bias for this datasets combination as the light blue line in Fig. \ref{fig:mass_bias}. The preferred bias value is $(1-b) = 0.67 \pm 0.04$, of the same order of the $\Lambda$CDM case. We highlight that when analysing these results we need to take into account the combined effect of different degeneracies between $(1-b)$, $\sum m_{\nu}$ and $\sigma _8$. In fact, the preferred high value of $\sigma _8$ from CMB primary anisotropies data still drives the constraints to lower values of the mass bias, despite the effect of massive neutrinos and the addition of tSZ probes, as shown in Fig. \ref{fig:mnu_ALL_params} right panel.
Finally, we find an upper limit on the neutrino mass of $\sum m_{\nu} < 0.23 \, \text{eV}$ at $95\%$ that is more stringent than constraints obtained from CMB alone ($\sum m_{\nu} < 0.49 \, \text{eV}$, from \cite{Ade:2015xua}).

\begin{table*}[!h]
\begin{center}
\scalebox{0.87}{
\begin{tabular}{c|c|c|c|c}
\hline
\hline
\horsp
 Cosmological parameters & $\text{NC}^{\text{tSZ}}$+ BAO &  $C_{\ell}^{\text{tSZ}} + \text{NC}^{\text{tSZ}}$+ BAO & CMB &CMB + $C_{\ell}^{\text{tSZ}} + \text{NC}^{\text{tSZ}}$+ BAO   \\
\hline
\morehorsp
 $\Omega _{\text{m}}$ & $ 0.337 _{-0.031}^{+0.027}$ & $ 0.335 _{-0.024}^{+0.023} $  & $ 0.353 _{-0.037}^{+0.020} $ & $0.315 \pm 0.008$\\
\hline
\morehorsp
$\sigma _8$ & $0.728 _{-0.038}^{+0.032} $ & $ 0.737 _{-0.037}^{+0.028} $ & $ 0.772_{-0.024}^{+0.049}$ & $0.792 _{-0.013}^{+0.020} $ \\
\hline
\morehorsp
$S_8 = \sigma_8 (\Omega _m/0.3)^{1/3}$ & $0.757 _{-0.040}^{+0.029} $  &$0.764_{-0.039}^{+0.028} $ & $ 0.813_{-0.024}^{+0.030} $& $0.804_{-0.015}^{+0.019} $ \\
\hline
\morehorsp
$\sum m_{\nu}$ & $ < 2.84 \, \text{eV}$ & $ < 1.88\, \text{eV}$  & $<0.68 \, \text{eV}$ &$ < 0.23 \, \text{eV}$\\
\hline
\hline
\morehorsp
 Nuisance parameters &$\text{NC}^{\text{tSZ}}$+ BAO & $C_{\ell}^{\text{tSZ}} + \text{NC}^{\text{tSZ}}$+ BAO & CMB  & CMB + $C_{\ell}^{\text{tSZ}} + \text{NC}^{\text{tSZ}}$+ BAO  \\
 \hline
\morehorsp
$(1-b)$ &$0.749 \pm 0.091 $ & $ 0.741 \pm 0.089 $  &  \bf - & $ 0.673 _{-0.047}^{+0.037}$ \\
\hline
\morehorsp
 $\alpha$ & $1.788 \pm 0.076 $ & $ 1.811_{-0.068}^{+0.077} $  &  \bf - & $ 1.824 _{-0.040}^{+0.037}$\\
\hline
\morehorsp
$\log Y_*$ & $-0.191 \pm 0.020 $ & $ -0.191_{-0.021}^{+0.023} $  &  \bf - & $-0.193 \pm 0.020 $\\
\hline
\morehorsp
 $\sigma _{\ln Y_*}$ &$0.075 \pm 0.010 $ & $ 0.075 \pm 0.010 $  & \bf - & $ 0.075 \pm 0.010 $\\
\hline 
\end{tabular}}
\caption{\footnotesize{$68\%$ c.l. constraints for cosmological and scaling-relation parameters and $95\%$ upper limits for neutrino mass for varying neutrino mass scenario, from number counts $(\text{NC}^{\text{tSZ}})$, the combination of power spectrum and number counts $(C_{\ell}^{\text{tSZ}}+\text{NC}^{\text{tSZ}})$, the addition of CMB data $(\text{CMB} \, + \, C_{\ell}^{\text{tSZ}}+\text{NC}^{\text{tSZ}}),$ and for CMB alone.}}
\label{tab:mnu}
\end{center}
\end{table*}

\subsubsection{Dark energy EoS}
\begin{figure}[!h]
\centering
\includegraphics[scale=0.33]{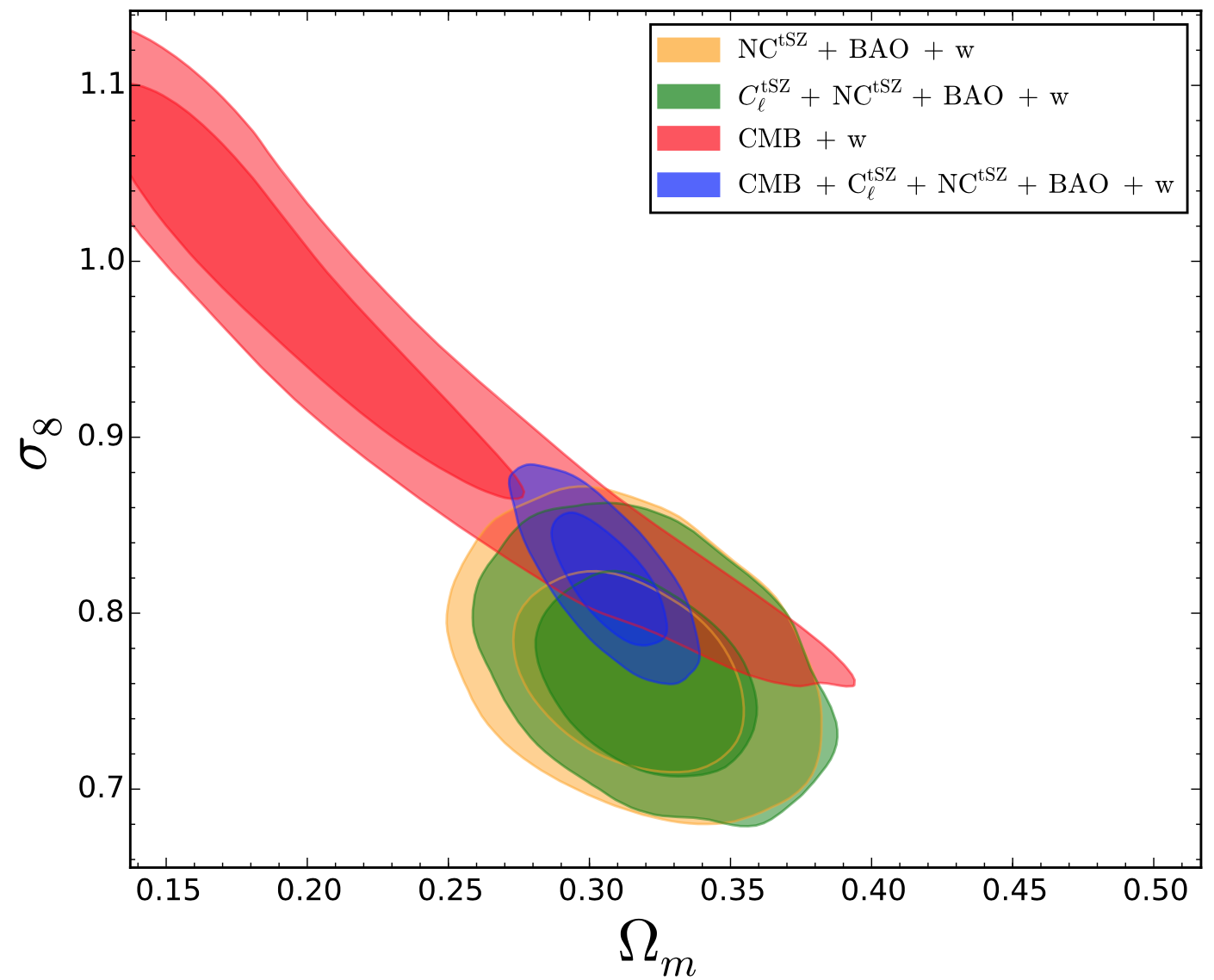}
\caption{\footnotesize{Two-dimensional probability distributions for $\Omega_m$ and $\sigma_8$ for varying dark energy EoS scenario, number counts (orange), the combination of number counts and power spectrum (green), the addition of CMB data (blue), and for only CMB (red).}}
\label{fig:DE}
\end{figure}

\begin{figure}[!h]
\centering
\includegraphics[scale=0.345]{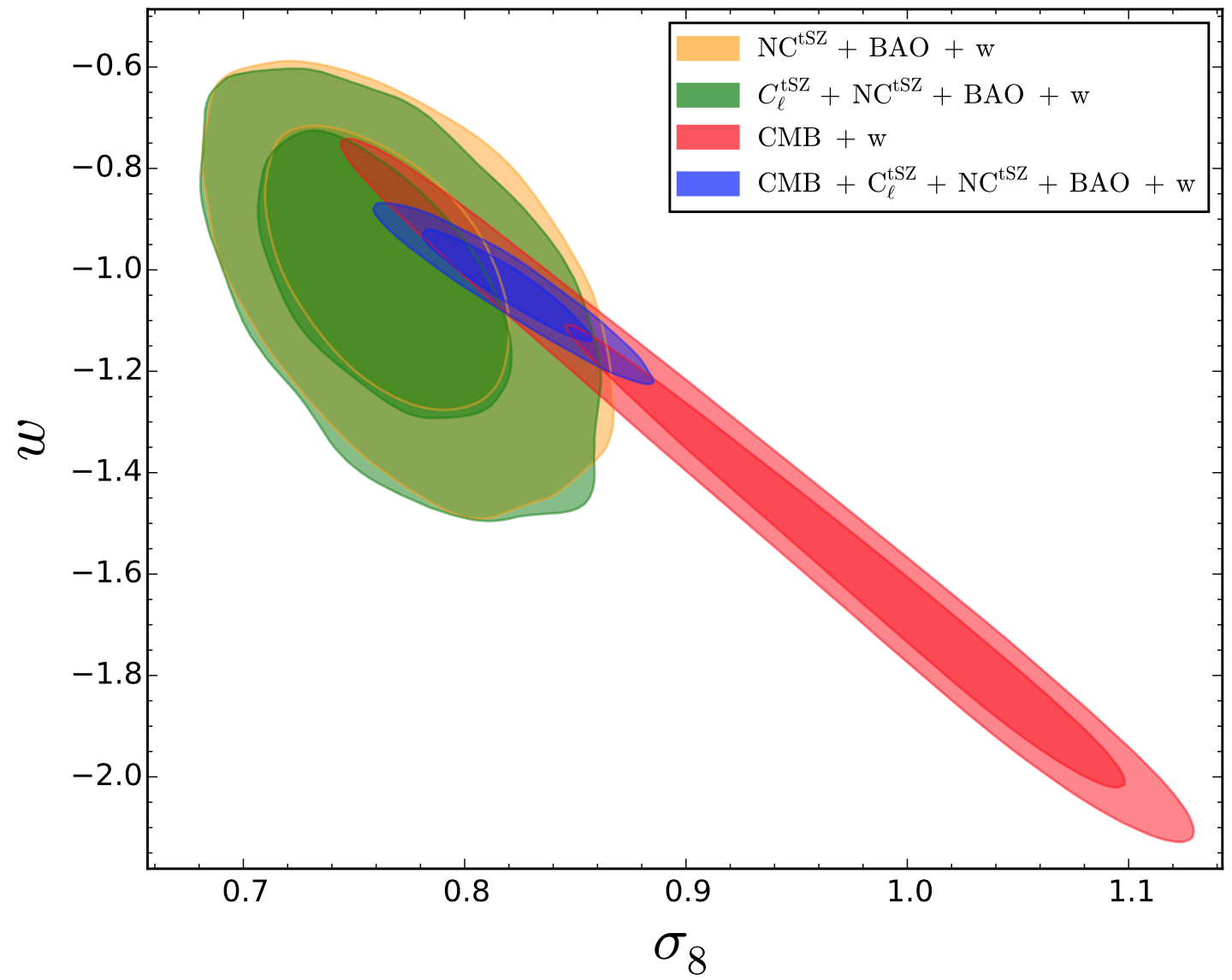}
\caption{\footnotesize{Two-dimensional probability distributions for $\sigma _8$ and $w$ for varying dark energy EoS scenario, for number counts (orange), the combination of number counts and power spectrum (green), the addition of CMB data (blue), and for only CMB (red).}}
\label{fig:DE_sigma8}
\end{figure}

We now consider the extension of the parameter space to dark energy EoS by allowing this parameter to differ from the standard value $w=-1$ for a cosmological constant. We focus on the simplest case, where $w$ is constant with time, to compare our results with those from \cite{Ade:2015fva}. We show the constraints on matter density $\Omega_m$ and $\sigma_8$ in Fig. \ref{fig:DE} and Tab. \ref{tab:de}. Again, CMB constraints are enlarged along a degeneracy line, but towards higher values of $\sigma_8$ and lower values of $\Omega_m$. Given this shift, we find an increased discrepancy between CMB and tSZ probes, at about $3.6 \, \sigma$, still driven by the $\sigma_8$ parameter, as shown in Fig. \ref{fig:DE_sigma8}. 
For the combination of tSZ probes we find a value of EoS parameter $w = -1.04 _{-0.17}^{+0.20}$ ($68\%$ c.l.), which is consistent with that found in \cite{Ade:2015fva} ($w = -1.01 \pm 0.18$ for number counts in combination with BAO). We underline that in this case the addition of tSZ power spectrum does not improve the results with respect to number counts alone, as shown in Figs. \ref{fig:DE} and \ref{fig:DE_sigma8}, while all results are reported in Tab. \ref{tab:de}.
For the complete combination of CMB and tSZ, we find the $68\%$ constraints $w= -1.03 _{-0.06}^{+0.08}$. We stress that these results (both for tSZ probes alone and in combination with CMB data) present a $1 \sigma$ consistency with the standard value $w=-1$, while results from CMB and BAO reported in \cite{Ade:2015fva} show only a $2 \sigma$ consistency. 

Finally, we stress that in this case as well the preferred value of the mass bias for the complete combination of CMB and tSZ probes is shifted to lower values, $(1-b) = 0.63 \pm 0.05$, as shown also in Fig. \ref{fig:mass_bias}, red line.

\begin{table*}[!h]
\begin{center}
\scalebox{0.87}{
\begin{tabular}{c|c|c|c|c}
\hline
\hline
\horsp
Cosmological parameters & $\text{NC}^{\text{tSZ}}$+ BAO & $C_{\ell}^{\text{tSZ}} + \text{NC}^{\text{tSZ}}$+ BAO & CMB & CMB + $C_{\ell}^{\text{tSZ}} + \text{NC}^{\text{tSZ}}$+ BAO \\
\hline
\morehorsp
 $\Omega _{\text{m}}$ & $0.315 _{-0.028}^{+0.025} $ & $ 0.321 _{-0.027}^{+0.024} $  & $ 0.209 _{-0.071}^{+0.023} $ & $ 0.306 \pm 0.013 $ \\
\hline
\morehorsp
$\sigma _8$ & $0.769 _{-0.041}^{+0.032} $ &$ 0.766 _{-0.042}^{+0.031} $  & $ 0.969 _{-0.057}^{+0.109} $ & $0.820 _{-0.027}^{+0.023} $\\
\hline
\morehorsp
$S_8 = \sigma_8 (\Omega _m/0.3)^{1/3}$ & $ 0.781_{-0.042}^{+0.030}$ & $ 0.782 _{-0.042}^{+0.031} $  & $0.846 \pm 0.020 $ &$0.826 \pm 0.018 $ \\
\hline
\morehorsp
$w$ & $ -1.01 _{-0.17}^{+0.20}$ &$ -1.04 _{-0.17}^{+0.20}  $ &$-1.56 _{-0.40}^{+0.21}$ &$ -1.03 _{-0.06}^{+0.08}$ \\
\hline
\hline
\morehorsp
 Nuisance parameters & $\text{NC}^{\text{tSZ}}$+ BAO &$C_{\ell}^{\text{tSZ}} + \text{NC}^{\text{tSZ}}$+ BAO &CMB & CMB + $C_{\ell}^{\text{tSZ}} + \text{NC}^{\text{tSZ}}$+ BAO  \\
 \hline
\morehorsp
$(1-b)$ & $0.750 \pm 0.091 $ & $0.751 \pm 0.092  $ &  \bf - & $ 0.634 _{-0.048}^{+0.040}$\\
\hline
\morehorsp
 $\alpha$ & $1.828 \pm 0.067 $ & $ 1.869 \pm 0.057 $  &  \bf - & $1.816 \pm 0.038$\\
\hline
\morehorsp
$\log Y_*$ & $-0.189 \pm 0.021 $ & $ -0.189 \pm 0.021 $  &  \bf - & $-0.194 \pm 0.020 $\\
\hline
\morehorsp
 $\sigma _{\ln Y_*}$ & $0.075 \pm 0.010 $ & $ 0.075 \pm 0.010 $  & \bf - & $ 0.075 \pm 0.010 $\\
\hline 
\end{tabular}}
\caption{\footnotesize{$68\%$ c.l. constraints for cosmological and scaling-relation parameters for varying dark energy EoS scenario, from number counts $(\text{NC}^{\text{tSZ}})$, the combination of power spectrum and number counts $(C_{\ell}^{\text{tSZ}}+\text{NC}^{\text{tSZ}})$, the addition of CMB data $(\text{CMB} \, + \, C_{\ell}^{\text{tSZ}}+\text{NC}^{\text{tSZ}}),$ and for CMB alone.}}
\label{tab:de}
\end{center}
\end{table*}

\section{Discussion}\label{sec:discussion}

\begin{figure}[H]
\centering
\resizebox{\hsize}{!}{\includegraphics{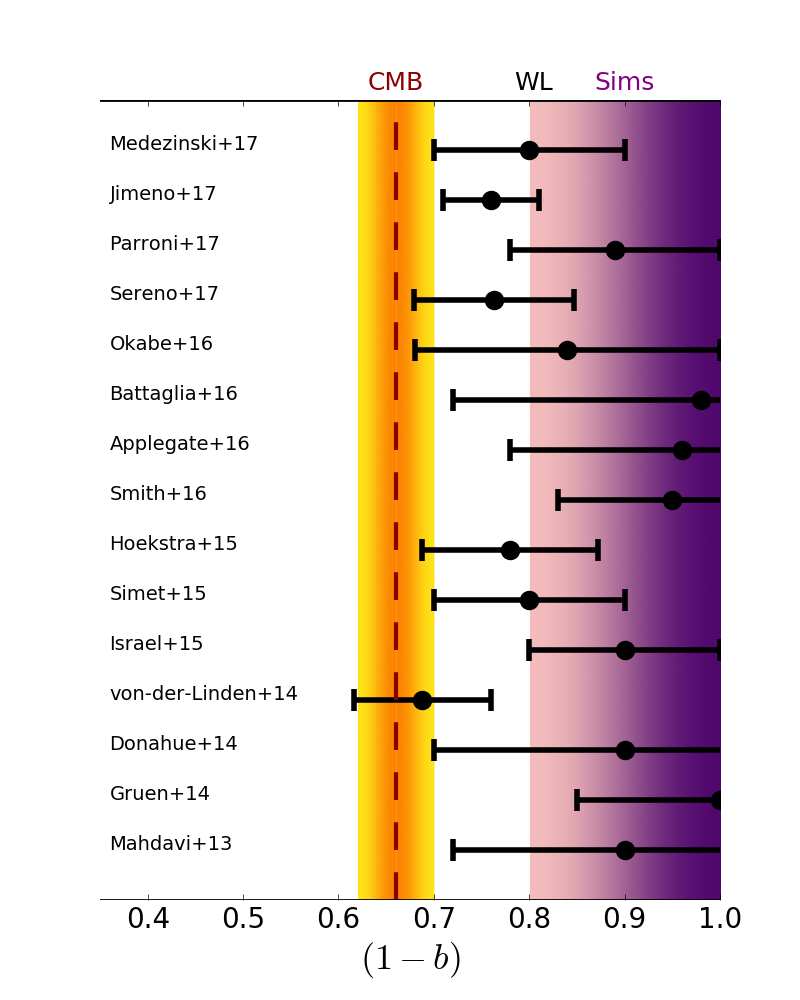}}
\caption{\footnotesize{Comparison of estimates of the mass bias from combined LCDM SZ and CMB analysis, WL-hydrostatic mass ratio, and simulations. See text for references to the mass bias estimates from the WL analyses.}}
\label{fig:bias_comp}
\end{figure}

The tSZ cluster counts and CMB tension reported in \cite{Ade:2015fva}, in agreement with constraints obtained independently from the tSZ power spectrum, one-dimensional pdf, and bispectrum \citep{2016A&A...594A..22P}, have triggered a lot of interest in the community. On the one hand, multiple estimates of the cluster masses were performed to investigate whether this discrepancy could be attributed to the mass bias (see Fig.~\ref{fig:bias_comp} for a summary of some of the most recent estimates). On the other hand, multiple cosmological analyses were performed to investigate the CMB/LSS tension or to try to reduce it. 

In this study, we provide constraints on cosmological parameters, considering as a baseline the updated value of the optical depth from \cite{Aghanim:2016yuo}, $\tau = 0.055 \pm 0.009$ and including the SPT high multipole tSZ spectrum estimate \citep{George:2014oba}. The new value of $\tau$ modifies the results from CMB primary anisotropies, increasing the constraining power on $\sigma _8$ of about $1 \sigma$. This is due to the dependence of the CMB power spectrum small-scale regime to the combination $\sigma _8 e^{-\tau}$. Results on $\sigma _8$ from number counts remain unchanged given that we use the approach and cluster sample from  \cite{Ade:2015xua}. For the tSZ effect, we assume the same baseline model as that of  \cite{Ade:2015fva}, i.e. we use a mass function from \cite{Tinker:2008ff} and a Gaussian prior on mass bias from \cite{Hoekstra:2015gda}, which is in agreement with the average mass bias obtained from the recent weak lensing (WL) estimates (Fig.~\ref{fig:bias_comp}). In this way, we can more easily and directly compare with the results from \cite{Ade:2015fva}.

The changes in the CMB results when considering the new value of $\tau$ reduce the discrepancy with tSZ counts, i.e. from $2.4 \, \sigma$ to $1.5 \, \sigma$ in the present study. We perform an actual combined analysis of tSZ number counts and power spectrum to carry out a complete MCMC exploration of the parameter space, sampling at the same time on cosmological and scaling-relation parameters. We neglect the correlation between the two combined probes in the likelihood as it is expected to be low with the current Planck cluster sample and large-scale power spectrum estimate \citep{Hurier:2017jgi}.  We find that the addition of tSZ power spectrum (including Planck and SPT) leads to $\sim 1.8 \, \sigma$ tension on $\sigma _8$ when compared to CMB results. Recent studies, using the LSS probes, also seem to show a disagreement with the best cosmology of the CMB. This includes studies based on cluster samples \citep{vik09,2013JCAP...07..008H,2013ApJ...763..147B,2016MPLA...3140008B,2017AJ....153..220B,Schellenberger2017}, on the linear growth rate data \citep[][and references therein]{Moresco2017}, or on cosmic shear \citep{2017MNRAS.465.1454H, 2017arXiv170605004V, Joudaki2017}. Despite intrinsic limitations to each of these probes (e.g. \cite{Efstathiou:2017rgv}), the LSS cosmological analyses exhibit a general trend towards lower values of $\sigma_8$.

\begin{figure}[!ht]
\centering
\resizebox{\hsize}{!}{\includegraphics{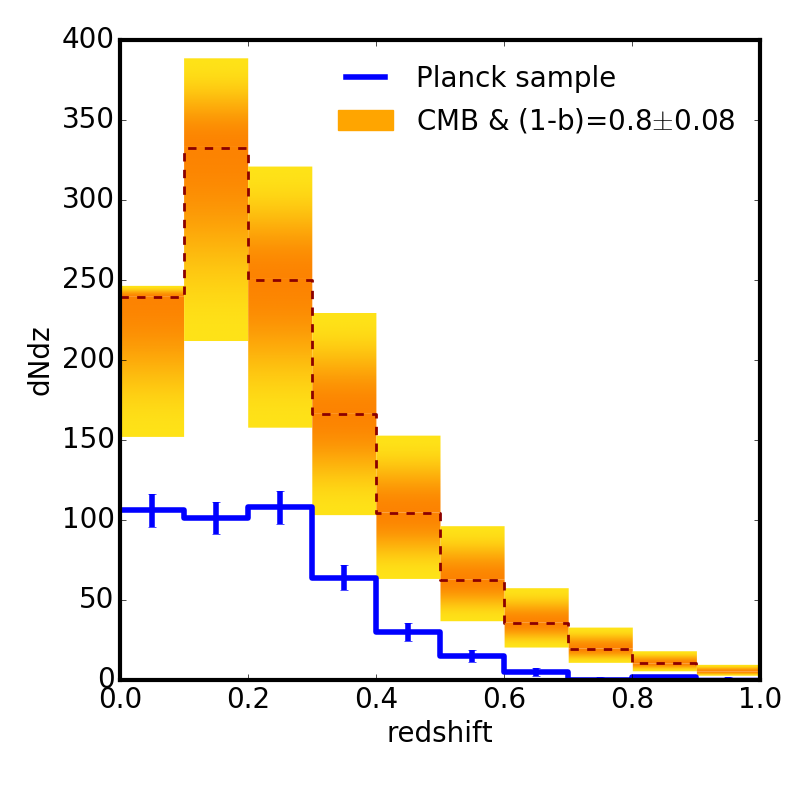}}
\caption{\footnotesize{tSZ cluster sample from Planck (blue line) compared with predicted counts with CMB best-fit cosmological parameters and $(1-b)=0.8\pm0.08$ (red line and orange envelope). }}
\label{fig:counts}
\end{figure}

\begin{figure}[!h]
\centering
\resizebox{\hsize}{!}{\includegraphics{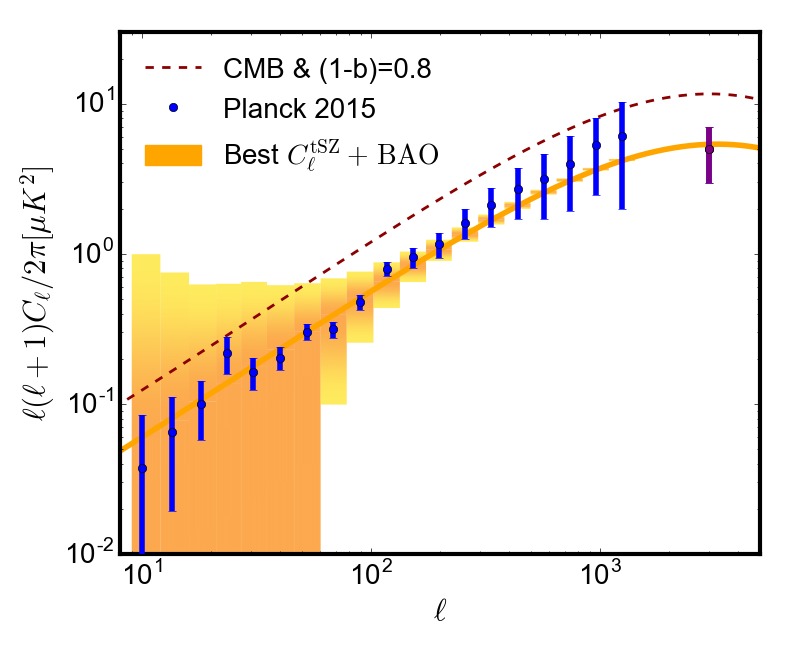}}
\caption{\footnotesize{tSZ power spectrum from Planck (blue points) and SPT (purple point) compared with predicted counts with CMB best-fit cosmological parameters and $(1-b)=0.8$ (red line) and our best fit (orange envelope).}}
\label{fig:spectrum}
\end{figure}

It was noticed in \cite{Ade:2015fva} that there was a factor $\sim 2.5$ more clusters predicted than observed when taking into account the CMB cosmology and a mass bias of 0.8. The new optical depth reduces the $\sigma _8$ derived from the CMB analysis to $\sigma _8 = 0.817 \pm 0.018$. Nevertheless, assuming a mass bias of 0.8 (average value of recent WL estimates), due to the high value of $\sigma _8$, we still find a difference between predicted and observed low redshift ($z<0.3$) cluster number counts of the order of 2.5 (Fig.~\ref{fig:counts}). We obtain a consistent discrepancy also for the tSZ power spectrum. Assuming again a mass bias of 0.8, the predicted power spectrum from CMB data shows an amplitude a factor of two higher than the measured tSZ power spectrum, as shown in Fig.~\ref{fig:spectrum}.

More biased estimates of the cluster mass could explain this difference, and in turn reduce the tSZ/CMB tension since  cosmological parameters are degenerate with scaling-relation parameters. We thus focus on the mass bias $(1-b)$. We show results for the combination of CMB primary anisotropies and number counts and the complete combination of CMB and tSZ probes, all using the updated value of the optical depth. For the first case, we find a low value for the mass bias $(1-b)=0.62 \pm 0.07$ that is fully consistent with results from \cite{Ade:2015fva}, while for the latter case, the addition of tSZ power spectrum data increases the bias of about $2 \sigma$ with respect to \cite{Ade:2015fva}. This leads to $(1-b)=0.63 \pm 0.04$. Hydrostatic mass estimates from X-ray observations (used to derive the scaling relation of Eq.~\ref{eq:Y500}) are known to be biased low from numerical simulation, but by not more than 20\% (\cite{Lau:2013ora,2016ApJ...827..112B}, and a
compilation of comparisons in \citet{2014A&A...571A..20P}, purple area in Fig.~\ref{fig:bias_comp}). With higher number of high-resolution optical observations of cluster samples and improved weak lensing mass measurements, many comparisons were made between X-ray or tSZ masses and WL masses. Assuming that WL reconstruction provides unbiased estimates of the true mass, many teams derived bias estimates  \citep[e.g.][shown as black dots, from top to bottom, in Fig.~\ref{fig:bias_comp}]{2017arXiv170600434M,Sereno:2017zcn,2017arXiv170600395J,Parroni:2017prn,2016MNRAS.461.3794O, 2016JCAP...08..013B,2016MNRAS.457.1522A,2016MNRAS.456L..74S,Hoekstra:2015gda,2015AAS...22544304S,2015MNRAS.448..814I,2014MNRAS.443.1973V,2014ApJ...794..136D,2014MNRAS.442.1507G,2013ApJ...767..116M}. These biases average around a value of $(1-b)\sim 0.8\pm0.08$;  one low mass bias was estimated in \cite{2014MNRAS.443.1973V}. 

As seen in Fig.~\ref{fig:bias_comp}, where we plot the highest mass bias possibly obtained from our combined analysis, we find values of $(1-b)$ that do not agree with those derived from numerical simulations and observations. Moreover, reducing the SZ-CMB tension by allowing lower values of $(1-b)$ does not seem to be sufficient for alleviating the global difference between low-$z$- and high-$z$-based cosmological parameters. 

Another way to reconcile CMB and tSZ is to relax some assumptions of the standard model. In particular, we find $(1-b)= 0.67 \pm 0.04 $ when adding massive neutrinos and $(1-b)= 0.63 \pm 0.04$ when opening the parameter space to a varying EoS parameter. This implies that even by exploring extension to $\Lambda$CDM, the complete combination of CMB and tSZ probes still points towards low values of $(1-b)$ and does not allow us to fully reconcile the probes.

As of today, the cluster number counts do not suffer from statistical uncertainty, but the counts are rather limited by systematic effects, mainly the mass estimates. The tSZ power spectrum, in turn, is not measured with sufficient accuracy, especially at small angular scales, to reduce the tension with CMB. The tSZ cosmological analysis can be improved by considering more realistic and complex  hypotheses on the mass bias (e.g. redshift and/or mass dependence), the pressure profile, and mass function. Regarding the latter, we recall here that the choice of mass function affects the final results on cosmological parameters, as shown, for example in \cite{Ade:2015fva}. For the current accuracy and precision obtained by tSZ probes measurements, different choices produce only a $\sim 1 \, \sigma$ shift of constraints along the same degeneracy line in the $(\Omega _m, \sigma_8)$ plane of parameters, therefore not affecting the discrepancy between tSZ probes and CMB data. Nevertheless, with future and more accurate measurements, it would also be necessary to marginalize over all the nuisance parameters (e.g. mass function fitting formula and bias dependencies).

The analysis can also be improved by refining the likelihood analysis, such as including correlations between probes, missing or inaccurate redshifts (e.g. \cite{Bonaldi:2014uza}), and  additional tSZ probes (e.g. bispectrum; \cite{Hurier:2017jgi}). 

\section{Conclusion}\label{sec:conclusions}

We have updated the constraints on cosmological parameters from tSZ cluster counts and power spectrum, using the most recent optical depth value from Planck and performing a combined analysis of the two probes; we have also added CMB data. In the $\Lambda$CDM case, we find that the combined analysis of tSZ counts and power spectrum improves the accuracy on $\Omega_m$ and $\sigma_8$ constraints slightly and  leads to a discrepancy of almost $1.8 \, \sigma$ on $\sigma _8$ when compared to CMB results. 

We then consider the effect of massive neutrinos, finding that the combination of tSZ counts and power spectrum allows us to obtain an upper limit on the neutrino mass, $\sum m_{\nu} < 1.88 \, \text{eV}$, resulting in an improvement of almost 30\% compared to number counts alone. Despite being weak  compared to other cosmological probes, tSZ data alone provide us with an independent constraint that can be combined in particular with CMB. In this case, the combination of tSZ probes and CMB leads a 95\% upper limit on massive neutrinos of $\sum m_{\nu} < 0.23 \, \text{eV}$. Moreover because of the enlargement of CMB constraints, we find that CMB results and combined tSZ results on $\sigma _8$ agree within $1.3 \, \sigma$. When we allow the EoS parameter $w$ for the dark energy to vary, the tSZ and CMB still show a higher $3.6 \, \sigma$ discrepancy on $\sigma _8$. 
 The full combination of probes provides $w =-1.03\pm 0.07$ that is consistent with the standard $w=-1$ value. 

Finally, we find that the complete combination of tSZ probes and CMB data points towards low values of the mass bias (almost $2 \, \sigma$ discrepancy) with respect to simulations and to other tracers of large-scale structure. Such a difference between mass estimates implies that, regardless of the CMB/LSS tension, a better understanding of the intrinsic systematic effects and differences between probes is needed.


\begin{acknowledgements}
      The authors acknowledge helpful discussions with Boris Bolliet. LS acknowledges support from the grant "Borsa di studio di perfezionamento all'estero" from University of Rome, Sapienza and "Fondazione Angelo della Riccia". Based on observations obtained with Planck (http://www.esa.int/Planck), an ESA science mission with instruments and contributions directly funded by ESA Member States, NASA, and Canada. This project has received funding from the European Research Council (ERC) under the European Union's Horizon 2020 research and innovation programme grant agreement ERC-2015-AdG 695561.
\end{acknowledgements}

\bibliographystyle{aa} 
\bibliography{references} 

\end{document}